\documentclass[twoside,11pt] {article}

\begin{document}

\title{Schouten-Nijenhuis Bracket of Killing-Yano Tensors for Kimura IIC Metric}

\author{Florian Catalin Popa \thanks{E-mail:~~~catalin@spacescience.ro}\\
{\small \it Institute of Space Science - Subsidiary of INFLPR}\\
{\small \it Atomistilor 409, RO 077125, P.O. Box MG 23}\\
{\small \it Magurele, Bucharest, Romania}}

\date{}
\maketitle

\begin{abstract}

We investigate Schouten-Nijenhuis brackets of Killing-Yano tensors for Kimura IIC metric showing that Killing-Yano tensors form a Lie algebra. 

\end{abstract}

\section{Introduction}

Killing-Yano tensors and their corresponding Killing tensors
have been studied extensively \cite{Carter,CMcL,Visi1,Visi2,Visi3, Klishevich} in the related context of finding solutions of the Dirac-equation in non-trivial curved space-time. Moreover, in the context of generalized Dirac-type operators \cite{tanimoto, cariglia, FCP} the Killing-Yano tensors are indispensable tools.
For the geodesic motion in the Taub-NUT space, the conserved vector analogous to the Runge-Lenz vector of the Kepler type problem is quadratic in $4$-velocities, its components are Killing tensors and they can be expressed as symmetrized products of Killing-Yano tensors \cite{Ya, Gibbons, GiRu1, GiRu2, VaVi1, vH}.
A non-degenerate closed conformal Killing–Yano 2-form, called the principal tensor, play an important role in the study of hidden symmetries in higher-dimensional
Kerr–NUT–(A)dS black holes. A comprehensive review on Killing-Yano tensors and their applications in black hole physics is given in \cite{FKK}.
The Schouten-Nijenhuis bracket plays an important structural role when applied to Killing–Yano tensors, especially in the geometry of symmetries of curved spacetimes and in the theory of integrable systems.
The antisymmetric Schouten-Nijenhuis brackets play an important role for the commutation of first-order symmetry operators commuting with the Dirac operator being important for the theory of separability of the Dirac equation. 
On the other hand, vanishing of the Schouten-Nijenhuis bracket of two Killing-Yano tensors has an important consequence for the theory of spinning particles and hence, vanishing of the Schouten-Nijenhuis bracket of two Killing-Yano tensors is a necessary requirement for two nongeneric supercharges to Dirac-Poisson-commute. 
Also, the authors of \cite{CKK} introduced a new bracked called Killing-Yano and demonstrate that it is closely related to the Schouten-Nijenhuis bracket.

It is well known that skew-symmetric tensors form a (graded) Lie algebra with respect to the skew-symmetric Schouten - Nijenhuis bracket \cite{Scho1940,Scho1954, Nij1955}. This fact led the authors of \cite{Kastor2007} to investigate whether, similar to Killing vectors, Killing-Yano tensors form a subalgebra of this algebra.  This statement is not true in general, the authors were able to give two counter examples disproving the conjecture. On the other hand, the statement is true in maximally symmetric spaces. 

The aim of this paper is to show that for a given  metric (Kimura IIC) the associated Killing-Yano tensor form a Lie algebra with respect to Schouten - Nijenhuis bracket. 
In the next two sections Killing and Killing-Yano tensors are revisited and their components are given for Kimura IIC metric. In the last section, after a short review, the Schouten - Nijenhuis bracket is evaluated. 

\section{Killing and Killing-Yano tensors}
There are two generalizations of the Killing vector in general relativity discussed extensively since the discovery of the Carter constant in the Kerr spacetime:

({\it a}) An antisymmetric tensor $f_{\mu_1 \mu_2... \mu_n}$ is
called a Killing-Yano tensor of rank $n$ if the following equation
is satisfied
\begin{equation}\label{ky}
f_{\mu_1 \mu_2... (\mu_n; \lambda)}=0.
\end{equation}

({\it b}) A symmetric tensor field $K_{\mu_1 \mu_2... \mu_n}$ is
said to be a Killing tensor of rank $n$ if:
\begin{equation}\label{k}
K_{(\mu_1 \mu_2...\mu_n; \lambda)}=0.
\end{equation}

Using these two generalizations, a
tensor of rank two may be constructed
$
K_{\mu \nu}=f_{\mu \mu_1 \mu_2... \mu_n}f^{\mu_1 \mu_2...
\mu_n}_{\ \ \ \ \ \ \ \ \ \nu}
$
which is a Killing tensor because is symmetric and satisfies the
Killing tensor equation (\ref{k}) with quantity $K_{\mu
\nu}p^{\mu}p^{\nu}=f_{\mu \mu_1 \mu_2... \mu_n}f^{\mu_1 \mu_2...
\mu_n}_{\ \ \ \ \ \ \ \ \ \nu}p^{\mu}p^{\nu}$ the quadratic first
integral generated by Killing tensor. In fact, a suitable construction  of Killing
tensor $K^{(ij)}_{\mu \nu}$ is given by
$
K^{(ij)}_{\mu \nu}=f^{(i)}_{ \mu_1 \mu_2... \mu_{n-1} (\mu}f^{(j)
\mu_1 \mu_2... \mu_{n-1}}_{\nu)}
$
where $f^{(i)}_{\mu \mu_1 \mu_2... \mu_n}$, indexed in what follows by
upper latin indices, represent the linear independent solutions for the Killing-Yano
equation (\ref{ky}). 

The metric tensor $g_{\mu \nu}$, as well as all
symmetrized products of any Killing vectors $\xi_I$, and in
general, a linear combination of them with constant coefficients,
are Killing tensors, i.e.
\begin{equation}\label{f0}
K_{\mu \nu}=a_0g_{\mu
\nu}+\sum\limits_{I=1}^N\sum\limits_{J=I}^Na_{IJ}\xi_{I(\mu}\xi_{|J|\nu)}
\end{equation}
we can say that all such Killing tensors are called $reducible$
and all other are called $ireducible$.

The rank one and four Killing-Yano tensors can
be considered as two trivial cases. In the first case, the
Killing-Yano tensor $f^{(i)}_{\mu}$ is a Killing vector and the
associated Killing tensor
$
K^{(ij)}_{\mu \nu}=f^{(i)}_{(\mu}f^{(j)}_{\nu)}
$
is obviously reducible. In the second case, the Killing-Yano
tensor $f^{(i)}_{\mu \nu \lambda \rho}$, solutions of the
equations $f^{(i)}_{\mu \nu \lambda (\rho; \theta)}=0$, is
proportional to the alternating pseudotensor $\eta_{\mu \nu
\lambda \rho}$. In this case the associated  Killing tensor
$
K^{(ij)}_{\mu\nu}=f^{(i)}_{\alpha\lambda\rho(\mu}f_{\nu)}^{(j)
\alpha\lambda\rho}
$
is equal to the metric up to a constant.

The Killing-Yano tensors $f^{(i)}_{\mu \nu}$ of rank two are
objects indispensable in the construction of new first integrals of the
geodesic motion, Dirac type operators and gravitational anomalies
investigations.

The
associated Killing tensors are given by
\begin{equation}\label{kt2}
K_{\mu\nu}^{(ij)} = f_{(\mu\vert{}\alpha\vert{}}^{(i)} f_{\nu)}^{(j)\alpha}
\end{equation}
some of which are expected to be irreducible.

In order to determine the Killing-Yano tensors
$f^{(i)}_{\mu\nu\lambda}$ of rank three a system with fifteen
independent equations for four independent components
$f^{(i)}_{\mu \nu (\lambda; \rho)}=0$ must to be solved. The
associated Killing tensors are given by
\begin{equation}\label{kt3}
K^{(ij)}_{\mu \nu}=f^{(i)}_{\alpha\beta(\mu
}f_{\nu)}^{(j)\alpha\beta}
\end{equation}
checking which of these are irreducible by comparison with the
definition of reducibility.

\section{Killing Tensors for Kimura IIC Metric }
Hauser and Malhiot \cite{Hauser-Malhiot} investigated static
spherically symmetric space-times and found all metrics for which
an irreducible Killing tensor of rank two exists obtaining
quadratic first integrals under the assumption that these are
independent of time. Without imposing such assumption Kimura
\cite{Kimura1} generalize the results obtained before by Hauser
and Malhiot and found several metrics having quadratic first
integrals. One of them is know in literature as the Kimura IIC
metric. This metric is of Petrov type $D$, with non-zero
energy-momentum tensor, given by
$$
ds^2=\frac{r^2}{b}dt^2-\frac1{r^2b^2}dr^2-r^2(d\theta^2+\sin^2\theta
d\phi^2)
$$
From Killing vector equations
$$
\xi_{(\mu;\nu)}=0,
$$
we find four Killing vectors
${\bf \xi}^{(i)}=\xi_{(i)}^{\mu}\partial_{\mu}$ given by
$$
\begin{array}{llll}
{\bf \xi}^{(0)}=\frac{\partial}{\partial t} & {\bf \xi}^{(1)}=\cos \phi \frac{\partial}{\partial \theta}-\cot \theta
\sin \phi \frac{\partial}{\partial \phi}\\
{\bf \xi}^{(2)}=-\sin \phi \frac{\partial}{\partial \theta}-\cot
\theta \cos \phi \frac{\partial}{\partial \phi} & {\bf \xi}^{(3)}=\frac{\partial}{\partial \phi}
\end{array}
$$
In covariant form the Killing vectors components reads
\begin{equation}\label{K1}
\begin{array}{llll}
{\bf\xi}_{(0)}=(\frac{r^2}{b}, 0, 0, 0)&{\bf\xi}_{(1)}=(0, 0, -r^2\cos \phi, r^2 \sin \theta \cos
\theta \sin \phi)\\
{\bf\xi}_{(2)}=(0, 0, r^2\sin \phi, r^2 \sin \theta \cos
\theta \cos \phi)&{\bf\xi}_{(3)}=(0, 0, 0, -r^2\sin^2 \theta)
\end{array}
\end{equation}

In what follows we construct Killing tensors from Killing-Yano
tensors of various rank and using (\ref{f0}) we check which of
them are irreducible.

For Killing-Yano tensors of rank two the solution of
corresponding equations (\ref{kt2}) is
\begin{equation}\label{KY2}
f_{23}=r^3\sin\theta,
\end{equation} 
which exist in any spherically symmetric static space-time
\cite{how, HC2}. The corresponding components of the Killing
tensor are
$
K_{22}=-r^4,\quad K_{33}=-r^4\sin^2\theta
$
and it is easy to observe that this tensor can be written as a linear
combination of symmetrized products of Killing vectors
$
K_{\mu\nu}=-\sum_{i=1}^{3}\xi_{i(\mu}\xi_{|i|\nu)},
$
which means
that this Killing tensor is reducible, identified with the square of the angular momentum.

In \cite{PT} was shown that irreducible Killing
tensors can be computed from third rank Killing-Yano tensors.
By straightforward calculation the following general
solution for Killing-Yano tensors of rank three is obtained,
$$
f_{023}=Ebtr^4\sin\theta+Fr^4\sin\theta,\quad f_{123}=Er\sin\theta
$$
all other components being zero. 
If the integration constants $E$ and $F$ equals with unity  two ($i=1,2$) independent solutions $f^{(i)}_{\mu\nu\lambda}$ are obtained,
\begin{equation}\label{sol1}
f^{(1)}_{023}=r^4\sin\theta
\end{equation}
and
\begin{equation}\label{sol2}
f^{(2)}_{023}=btr^4\sin\theta,\quad f^{(2)}_{123}=r\sin\theta
\end{equation}

Taking into account (\ref{kt3}) the following
Killing tensors from the Killing-Yano tensors (\ref{sol1}) and
(\ref{sol2}) can be constructed.

$\bullet$ {\it Case 1}: If only the solution (\ref{sol1}) is considered then the Killing tensor $K^{(11)}_{\mu\nu}$ haves the following
non-null components,
\begin{equation}\label{corrKcase1}
K^{(11)}_{00}=2r^4,\quad K^{(11)}_{22}=-2br^4,\quad K^{(11)}_{33}=-2br^4\sin^2\theta
\end{equation}
which is reducible since can be written as a linear combination of
symmetrized products of Killing vectors
$$
K^{(11)}_{\mu\nu}=2b^2\xi_{0(\mu}\xi_{|0|\nu)}-2b\sum_{i=1}^{3}\xi_{i(\mu}\xi_{|i|\nu)}.
$$

$\bullet$ {\it Case 2}: If only the solution (\ref{sol2}) is considered then the Killing tensor $K^{(22)}_{\mu\nu}$ haves the following
non-null components,
$$
K^{(22)}_{00}=2b^2t^2r^4,\quad K^{(22)}_{01}=2btr,\quad K^{(22)}_{11}=\frac{2}{r^2},
$$
\begin{equation}\label{corrKcase2}
K^{(22)}_{22}=2b^2r^2(1-bt^2r^2),\quad K^{(22)}_{33}=2b^2r^2(1-bt^2r^2)\sin^2\theta.
\end{equation}
If a new Killing tensor is defined $K^{\prime(22)}_{\mu\nu}\equiv
K^{(22)}_{\mu\nu}+ag_{\mu\nu}$, for $a=2b^2$, we obtain 
$$
K^{\prime(22)}_{00}=2br^2(bt^2r^2+1),\quad K^{\prime(22)}_{01}=2btr
$$
\begin{equation}\label{corrKpr}
K^{\prime(22)}_{22}=-2b^3t^2r^4,\quad K^{\prime(22)}_{33}=-2b^3t^2r^4\sin^2\theta
\end{equation}
which is irreducible since it is clearly impossible to obtain from
metric and Killing vectors terms which are explicit functions of
$t$.

$\bullet$ {\it Case 3}: Considering both solution (\ref{sol1}) and
(\ref{sol2}) of the third rank Killing-Yano equation, then
the following Killing tensor is obtained, with components,

\begin{equation}\label{mixt}
K^{(12)}_{00}=2btr^4,\quad K^{(12)}_{01}=r,\quad K^{(12)}_{22}=-2b^2tr^4,\quad K^{(12)}_{33}=-2b^2tr^4\sin^2\theta
\end{equation}
We observe that this tensor can not be obtained as linear
combination of the metric and symmetrized products of Killing
vectors and is also irreducible.

For Killing-Yano tensors of rank four the solution of
corresponding equation (\ref{ky}) yields a system of 4 equations
with one $(f_{0123})$ independent component
$
f_{0123}=r^2\sin\theta.
$
We observe that for a four rank Killing-Yano tensor, which is
antisymmetric in all indices, the Killing-Yano equations are
equivalent with $f_{0123;\mu}=0$, and the corresponding Killing
tensor is reducible, proportional with the metric tensor.

To avoid sign ambiguities associated with the Lorentzian signature, we briefly specify our conventions for the Hodge dual. We adopt the metric signature $(+,-,-,-)$ and define the completely antisymmetric Levi-Civita tensor $\epsilon_{\mu\nu\rho\sigma}$ with the positive orientation $\epsilon_{0123} = +\sqrt{-g}$ (where $g$ is the determinant of the metric). Consequently, the covariant components of the dual vector field ($\star f$) generated from a totally antisymmetric rank-3 tensor $f_{\nu\rho\sigma}$ are evaluated as:$$(\star f)_\mu = \frac{1}{3!} \epsilon_{\mu\nu\rho\sigma} f^{\nu\rho\sigma}$$Using this standard convention, it is worth noting that the Hodge dual of a third rank Killing-Yano tensor is a closed conformal Killing vector. In our case, if we consider the Hodge dual of (\ref{sol1}) and (\ref{sol2}) we obtain the following conformal Killing vectors,
$$
{f}^{\ast (1)}=b\sqrt {b} r^2 \frac{\partial}{\partial r}
$$
$$
{f}^{\ast (2)}=-\frac{b \sqrt {b}}{r} \frac{\partial}{\partial
t} + b^2\sqrt {b}tr^2 \frac{\partial}{\partial r}
$$
We observe that ${f}^{\ast}_{ (1)}$ and ${f}^{\ast}_{ (2)}$ are of gradient type, ${f}^{\ast}_{(1) \lambda}=(-r/{\sqrt {b}})_{,\lambda}$ and ${f}^{\ast}_{(2) \lambda}= (-\sqrt {b}tr)
_{,\lambda}$ with associated conformal factors $\vartheta ^{(1)} = b \sqrt{b}r$ and $\vartheta ^{(2)} = b^2
\sqrt{b}tr$.
\\

\section{Schouten--Nijenhuis Bracket and Killing-Yano Lie Algebra}

Given $A$ a rank $p$ multivector and $B$ a rank $q$ multivector, their Schouten--Nijenhuis bracket is a $(p+q-1)$-multivector defined by

\begin{equation}
[A,B]^{a_1\dots a_{p+q-1}}=pA^{b[a_1\dots a_{p-1}}\nabla_b B^{a_p\dots a_{p+q-1}]}
+ (-1)^{pq} qB^{b[a_1\dots a_{q-1}}\nabla_b A^{a_q\dots a_{p+q-1}]}\, . \label{SN_def}
\end{equation}
In \cite{Kastor2007} was shown that if $A^a$ and $B^a$ are two Killing vectors, their bracket $C^a=A^b\nabla_bB^a-B^b\nabla_bA^a$ satisfies Killing's equation
$\nabla_aC_b=\nabla_{[a}C_{b]}$. Also, the Schouten--Nijenhuis bracket of a Killing vector $A^a$ with a Killing-Yano tensor $B^{ab}$ is again a Killing-Yano tensor. For $C^{ab}=[A,B]^{ab}$, which in this case is the Lie derivative of $B^{ab}$ with respect to the Killing vector $A^a$, for the expression 
$C^{ab}=A^c\nabla_c B^{ab}- B^{cb}\nabla_cA^a-B^{ac}\nabla_c A^b$, was shown that $\nabla_aC_{bc} =\nabla_{[a}C_{bc]} $ and that therefore $C_{ab}$ is again a Killing-Yano tensor.
If $A^{ab}$ and $B^{ab}$ are two rank two Killing-Yano tensors, their SN bracket $C^{abc}=[A,B]^{abc}$ given by 
$C^{abc} = 2( A^{d[a}\nabla_dB^{bc]} + B^{d[a}\nabla_dA^{bc]} )$ it is not generally a Killing-Yano tensor.

In what follows, for Kimura IIC metric, we evaluate the Schouten--Nijenhuis bracket, where the Killing vectors are given in covariant form by (\ref{K1}), rank two Killing-Yano tensor by (\ref{KY2}) and third rank Killing-Yano tensors are given by (\ref{sol1}) and (\ref{sol2}) respectively.

The Schouten--Nijenhuis brackets, different from zero, are obtained for the following situations:

If $A$ and $B$ are Killing vectors (\ref{K1}) their Schouten--Nijenhus brackets reads:
$Case 1:$
Let $A = \xi_{(3)}$ and $B = \xi_{(1)}$ with
$$
A_{3}=-r^2\sin^2(\theta),\quad 
B_{2}=-r^2\cos(\phi),\quad 
B_{3}=r^2 \sin(\theta)\cos(\theta)\sin (\phi),
$$
then
$$
C_{0}=C_{1}=0,\quad 
C_{2}=r^2\sin(\phi),\quad 
C_{3}=r^2 \sin(\theta)\cos(\theta)\cos(\phi)
$$
The bracket evaluates to components which exactly correspond to the basis vector $C = \xi_{(2)}$.
\\
$Case 2:$
Let $A = \xi_{(3)}$ and $B = \xi_{(2)}$ with
$$
A_{3}=-r^2\sin^2(\theta),\quad 
B_{2}=r^2\sin(\phi),\quad 
B_{3}=r^2 \sin(\theta)\cos(\theta)\cos(\phi),
$$
then
$$
C_{0}=C_{1}=0,\quad 
C_{2}=r^2\cos(\phi),\quad 
C_{3}=-r^2 \sin(\theta)\cos(\theta)\sin(\phi)
$$
which exactly correspond to the basis vector $C = -\xi_{(1)}$.
\\
$Case 3:$
Let $A = \xi_{(1)}$ and $B = \xi_{(2)}$ with
$$
A_{2}=-r^2\cos(\phi),\quad 
A_{3}=r^2 \sin(\theta)\cos(\theta)\sin (\phi),\quad 
$$
and
$$
B_{2}=r^2\sin(\phi),\quad
B_{3}=r^2 \sin(\theta)\cos(\theta)\cos(\phi),\quad 
$$
then
$$
C_{0}=C_{1}=C_{2}=0,\quad 
C_{3}=-r^2\sin^2(\theta).
$$
which exactly correspond to the basis vector $C = \xi_{(3)}$.

The only nonzero Schouten--Nijenhuis brackets are obtained for the situation when $A= \xi_{(0)}$ is the Killing vector and $B = f^{(2)}$ is the rank three Killing-Yano tensor, with
$$
A_{0}= \frac{r^2}{b},\quad 
B_{023}=btr^{4}\sin(\theta),\quad 
B_{123}=r \sin(\theta),\quad 
$$
then
$$
C_{012}=C_{013}=C_{123}=0,\quad 
C_{023}=br^4\sin(\theta)
$$
and we observe that $C_{023}=bf_{023}^{(1)}$, corresponding exactly to the basis tensor $C = b f^{(1)}$.
We mention that for all other rank combinations associated to $A$ and $B$ their Schouten--Nijenhuis brackets vanishes.

A brief analysis shows that the three spatial Killing vectors represent the spherical symmetry of the metric and completely decouple to form the standard orthogonal Lie algebra $so(3)$. The time translation Killing vector and the two third rank Killing-Yano tensors form the nilpotent 3-dimensional Heisenberg algebra $h_3$. The remaining second rank Killing-Yano tensor and the forth rank Killing-Yano tensor, because they commute with absolutely everything in the space, they form a purely central, non-interacting abelian ideal.

As a final note, given that in four-dimensional spacetimes there are very few examples of irreducible Killing tensors of order higher than two \cite{GHKW}, for the present case, no higher-rank Killing tensors could be obtained, the symmetric Schouten-Nijenhuis brackets of the second-rank Killing tensors being always zero.

\section{Conclusions}
In this paper, for the Kimura IIC metric, we were able to prove by straightforward computation that the Killing and Killing-Yano tensors form a closed Lie superalgebra under the Schouten-Nijenhuis bracket. As expected for a spherically symmetric spacetime, the spatial Killing vectors close to form the standard $so(3)$ Lie algebra. Furthermore, when integrating the time-translation vector and the higher-rank Killing-Yano tensors, we find that the complete extended algebra perfectly decouples. Because the higher-rank tensors are constructed using the invariant spherical area form, they trivially commute with the spatial rotations. Consequently, the full geometric structure forms a direct sum $g \cong so(3) \oplus h_3 \oplus R^{0|2}$, containing the $so(3)$ isometry algebra, a 3-dimensional Heisenberg subalgebra $h_3$ generated by the time-translation and third rank tensors, and a central abelian ideal. 

\section*{Acknowledgments}
This research was supported by Romanian Ministry of Research,
Innovation and Digitalization under Romanian National Core Program LAPLAS VII – contract no. 30N/2023.

\end{document}